\documentclass[prb,twocolumn,showpacs,aps,superscriptaddress,floatfix]{revtex4-2}
\usepackage{amsmath}
\usepackage[normalem]{ulem}
\usepackage{graphicx}
\usepackage[dvipsnames]{xcolor} %provides more colours than "color" package
\begin{document}

\title{
%Temperature dependence of the anisotropy in the nematic superconductor as a probe of symmetry breaking mechanism%
Breaking of Ginzburg-Landau description in the temperature dependence of the anisotropy in the nematic superconductor
}
\author{M.I. Bannikov}
\affiliation{P.N. Lebedev Physical Institute, Russian Academy of Sciences, Moscow 119991, Russia}
\affiliation{National Research University Higher School of Economics, Moscow 101000, Russia}

\author{R.S. Akzyanov}
\affiliation{Dukhov Research Institute of Automatics, Moscow, 127055 Russia }
\affiliation{Moscow Institute of Physics and Technology, Dolgoprudny,
    Moscow Region, 141700 Russia}
\affiliation{Institute for Theoretical and Applied Electrodynamics, Russian
Academy of Sciences, Moscow, 125412 Russia}

\author{N.K. Zhurbina}
\affiliation{P.N. Lebedev Physical Institute, Russian Academy of Sciences, Moscow 119991, Russia}

\author{S.I. Khaldeev}
\affiliation{P.L. Kapitza Institute of Physical Problems, Moscow, Russia}
\affiliation{National Research University Higher School of Economics, Moscow 101000, Russia}

\author{Yu.G. Selivanov}
\affiliation{P.N. Lebedev Physical Institute, Russian Academy of Sciences, Moscow 119991, Russia}

\author{V.V. Zavyalov}
\affiliation{P.L. Kapitza Institute of Physical Problems, Moscow, Russia}
\affiliation{National Research University Higher School of Economics, Moscow 101000, Russia}

\author{A. L. Rakhmanov}
\affiliation{Dukhov Research Institute of Automatics, Moscow, 127055 Russia }
\affiliation{Institute for Theoretical and Applied Electrodynamics, Russian
Academy of Sciences, Moscow, 125412 Russia}

\author{A.Yu. Kuntsevich}
\affiliation{P.N. Lebedev Physical Institute, Russian Academy of Sciences, Moscow 119991, Russia}
\affiliation{National Research University Higher School of Economics, Moscow 101000, Russia}

\begin{abstract}
Nematic superconductors are characterized by an apparent crystal symmetry breaking that results in the anisotropy of the in-plane upper critical magnetic field $H_{c2}$. The symmetry breaking is usually attributed to the strain of the crystal lattice. The nature and the value of the strain are debatable. We perform systematic measurements of the $H_{c2}$ anisotropy in the high-quality Sr$_x$Bi$_2$Se$_3$ single crystals in the temperature range 1.8~K$<T<T_c\approx 2.7$~K using temperature stabilization with an accuracy of 0.0001 K. We observe that in all tested samples the anisotropy {is weakly temperaure dependent} when $T<0.8\,T_c$ and smoothly decreases at higher temperatures without any sign of singularity when $T\rightarrow T_c$. Such a behavior {is in a drastic contradiction with the prediction of} the Ginzburg-Landau theory for the nematic superconductors. {We discuss possible reasons for this discrepancy.}

\end{abstract}
\maketitle

Nematic superconductivity has been recently explored in different systems such as doped 3D topological insulators Bi$_2$Se$_3$~\cite{ Matano2016, Pan2016, Asaba2017}, PbTaSe$_2$~\cite{Le2020}, ({PbSe})$_{5}$(Bi$_{2}$Se$_{3}$)$_{6}$~\cite{Andersen2018}, few-layer NbSe$_2$~\cite{NbSe2}, Kagome metal CsV$_3$Sb$_5$~\cite{kagome}, {Bi$_2$Te$_3$/FeTe$_{0.55}$Se$_{045}$  heterostructures}~\cite{chen2018} and magic-angle twisted bilayer graphene~\cite{Cao2020}. In the nematic state the three-fold crystalline symmetry of the material is spontaneously broken and a single anisotropy axis arises. Along this axis the in-plane critical magnetic field $H_{c2}$ has a maximum while in the perpendicular direction $H_{c2}$ is minimal. The nematicity is experimentally observed not only in the transport and magnetic properties but also in specific heat measurements~\cite{Yonezawa2016,Sun2019}, nuclear magnetic resonance~\cite{Matano2016}, scanning-tunneling spectroscopy~\cite{Tao2018}, and thermal expansion~\cite{Cho2020}. From the theoretical point of view, the nematic superconductivity is a consequence of the vector nature of order parameter, that can be aligned in a definite direction by some symmetry-breaking field. The origin of the nematicity is now intensively debated in literature~\cite{Cho2020, Kostylev2020, Kawai2020}. 

For archetypical nematic superconductors of doped Bi$_2$Se$_3$ family different mechanisms are suggested to explain the nematicity. One possibility is {either} 
%{static asymmetry}: 
an initial lattice distortion acquired
during the crystal growth~\cite{Kuntsevich2018,KuntsevichPRB2019} or a structural stripeness in distribution of dopant atoms~\cite{Pan2016}. {The distortions arising due to such reasons exist in the sample in a wide temperature range up to room temperature and will be further referred to as external one.} %Observations of the 
Small deviations from the trigonal symmetry observed by high-resolution X-ray diffraction and nematic normal-state magnetoresistance well above $T_c$ could be considered as {confirmations of these mechanisms}~\cite{KuntsevichPRB2019}. The second possibility is a spontaneous formation of the nematicity. 
In Ref.~\onlinecite{kapranov} the existence of the spontaneous uniaxial strain in the superconducting state in the nematic superconductors {due to coupling of the superconducting order parameter with elastic deformation was predicted theoretically. This strain turns to zero near $T=T_c$. The existence of the spontaneous strain is consistent with the recent observations of the rotational symmetry breaking of the specific heat in the in-plane magnetic field in Sr$_x$Bi$_2$Se$_3$ in the superconducting state and just above the $T_c$ (possibly due to superconducting fluctuations)~\cite{Sun2019}. Similar observation of the nematicity in the thermal expansion was reported in Ref.~\onlinecite{Cho2020}. Note that neutron scattering experiments did not find any signature of a structural phase transition between 295 and 2~K in Cu$_x$Bi$_2$Se$_3$ but only small distortions of Bi atoms~\cite{Frolich2020}.
We discuss only single-domain crystals since in the multi-domain ones the applied uniaxial strain does not drive the nematicity direction but mainly affects the degree of the nematic anisotropy and aligns different nematic domains}~\cite{Kostylev2020}.

\begin{figure*}[t]
\center{\includegraphics[width=0.97\linewidth]{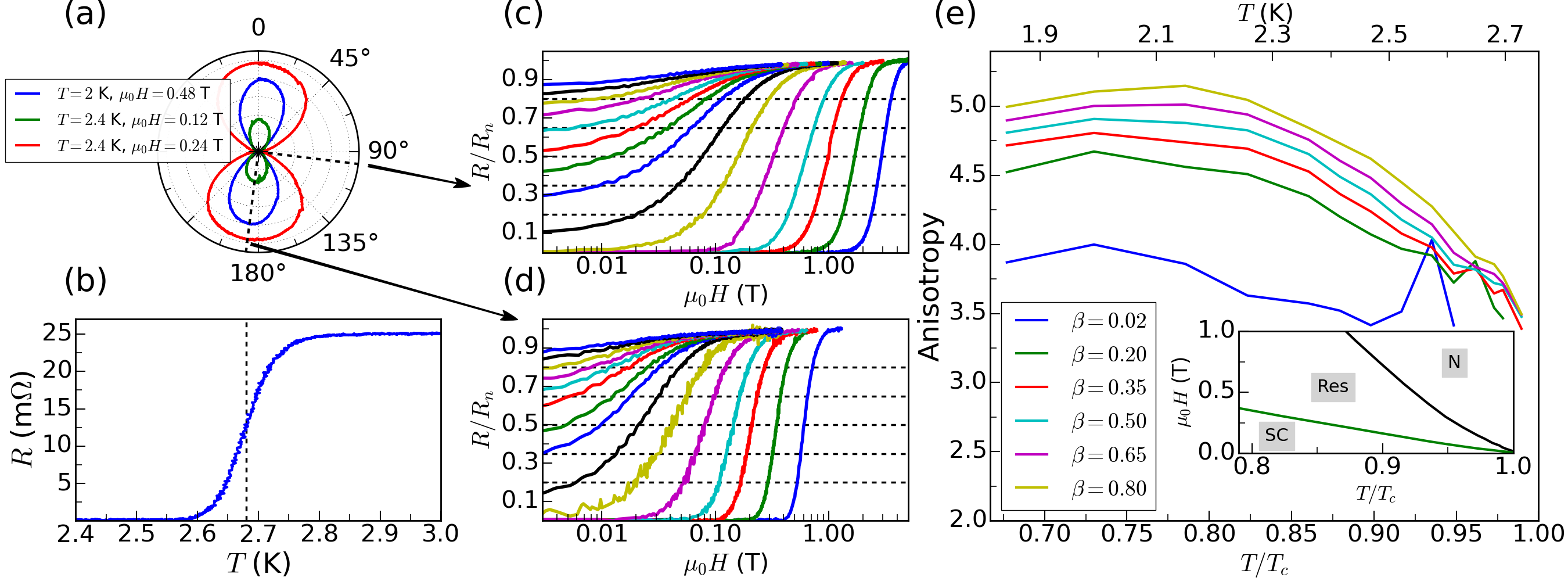}}
\caption{Experimental results {for the sample {\#317-c} 
Sr$_{0.15}$Bi$_2$Se$_3$.} a) Resistivity below $T_c$ vs polar angle for {several values of magnetic field and temperature;} 
{dashed} %dotted
lines indicate max- and min-$H_{c2}$ directions. b) Superconducting transition at $H=0$. Vertical {dashed} %dotted
line shows $T_c$ determined using 50\% criterion; c), and d) normalized $R(H)$ dependencies for max-$H_{c2}$ and min-$H_{c2}$ field orientations, respectively, for several temperatures (from right to left): 1.8 K, 2.1 K, 2.3 K, 2.4 K, 2.5 K, 2.57 K, 2.615 K, 2.645 K, 2.657 K, 2.67 K, 2.68 K, 2.69 K, 2.7 K, 2.71 K, 2.72 K. {e) the dependence of the anisotropy of $H_{c2}$ on temperature calculated using different criteria {$\beta$ indicated in panel}, the temperature normalized on $T_c$ determined by 50\% criterion. {Inset on panel (e) shows max-$H_{c2}$ and min-$H_{c2}$ dependencies (black and green lines respectively) on $T/T_c$ ($T_c$ determined by 50\% criterion). ``N'' stands for normal phase, ``Res'' stands for resistive phase with strong nematic anisotropy, ``SC'' stands for superconducting phase.}}}
%%%%%%%%%%%%%%%%%%%%%%%%%%%%%%%%%%%%%%%%%%%%%%%%%%%%%%%%%%
\label{Fig::results}
%%%%%%%%%%%%%%%%%%%%%%%%%%%%%%%%%%%%%%%%%%%%%%%%%%%%%%%%%%%
\end{figure*}

In this paper we report the results of measurements of the anisotropy {of the in-plane upper critical magnetic field $H_{c2}$ in the temperature range from $T=1.8$~K to $T_c\approx 2.7$~K for several high-quality (single-domain)} single crystals of doped topological insulator Sr$_x$Bi$_2$Se$_3$. We approach the critical temperature with the high accuracy of the temperature stabilization. We observe that the anisotropy of the 
{upper critical field} %second critical field typo(iii)
$An(T)$ {is weakly temperature-dependent  for 
$T<0.8 T_c$} %~K typo (i)
and decreases monotonically at higher temperatures. The drop of the anisotropy of $H_{c2}$ with $T$ is smooth without any sign of a singularity at $T\rightarrow T_c$ and the anisotropy remains significant (about
{$2$--$4$}) %$2 \div 4$ typo (ii) 
when $T=T_c$.   

{The obtained experimental results on the temperature dependence of the anisotropy is inconsistent with predictions of the conventional Ginzburg-Landau (GL) theory for nematic superconductors with two-component vector order parameter~\cite{Venderbos2016}. We can not explain them in the framework of the GL theory taking into account either external or spontaneous strain. Our results raise a question on the limits of applicability of the GL approach for a quantitative description of the nematic superconductivity in doped topological insulators.} 

{\bf Samples.} {In the superconductive doped bismuth selenides family, the Sr$_x$Bi$_2$Se$_3$ is of the highest structural quality}, uniformity, and temporal stability. We have grown high-quality crystals of Sr$_x$Bi$_{2}$Se$_3$ with
{$x=0.06$--$0.15$} %$x=0.1\div 0.15$ typo(ii)
using Bridgeman method~\cite{Kuntsevich2018}. Superconducting properties of these materials are weakly dependent on doping level $x$~\cite{Liu2015, Amoalem2020, KuntsevichPRB2019, Kuntsevich2019}. 
{Our crystals have a perfect morphology, high cristalline quality~\cite{Kuntsevich2018,KuntsevichPRB2019}, and almost 100 \% superconducting volume fraction~\cite{Amoalem2020}. The crystals of doped Bi$_2$Se$_3$ usually consist of several single-crystal domains~\cite{Kuntsevich2018,Tao2018,Kostylev2020,Du2017}.}
%Our crystals have a perfect morphology, and almost 100 \% superconducting volume fraction~\cite{Amoalem2020}. The X-ray diffraction have proven their high quality~\cite{Kuntsevich2018,KuntsevichPRB2019}. 
In the superconducting state such domains may have different nematicity orientation and even slightly different critical temperature. 
An accurate study of the effects near $T_c$ requires a pure single-domain selection. We cleaved such single blocks of typical sizes  
{1--2}~mm %1$\div$2 typo(ii)
in the $ab$ crystal plane and 0.1~mm along the $c$-axis from the same crystals that were used in our previous work~\cite{KuntsevichPRB2019}. 

{\bf Experimental.} The samples have been placed on the insert to the cryostat so that the $c$-axis of the crystal was perpendicular to the plane of rotation of the magnetic field {with 2$^\circ$ precision. It was checked previously that such inclination does not affect nematic superconducting properties in Sr$_x$Bi$_2$Se$_3$\cite{Kuntsevich2018, Kostylev2020}}. We denote the %rotation 
angle {between the current flow direction and } 
%of 
the applied magnetic field as $\theta$. Several contacts were glued to the samples {from different sides} to perform the measurements using standard lock-in technique%{, as shown in Fig.\ref{Fig::scheme}}
. The presence of several contacts allows us to vary the direction of the {in-basal-plane} transport current with respect to the crystal axes. We observe that the angular behavior of the in-plane upper critical magnetic field $H_{c2}$ is independent of the direction of the transport current, in agreement with the previous experiments~\cite{Kuntsevich2018,Yonezawa2018,KuntsevichPRB2019,Smylie2018}.
 
We extract the field $H_{c2}$ from the value of the magnetoresistance $R(H)$ measured at fixed temperature. Rotating the sample, we obtain the angular dependence of the magnetoresistance and $H_{c2}(\theta)$. The anisotropy $An$ is determined as the ratio of the highest value of $H_{c2}(\theta)$ to the lowest one. Near $T_c$ the upper critical field becomes small and measuring anisotropy with high precision requires a high stabilization of temperature, which was of the order of a few mK or less in our experiments.

Conventional cryogenic experimental systems provide a fast and convenient way to perform magnetotransport measurements at low temperatures with in-situ sample rotation. In such systems the sample is usually cooled by the vaporised $^4$He flow at low pressure. The thermal contact between the sample and the temperature sensor is usually far from perfect. The platform holding the sample rotates with respect to the field and not vice versa. In different positions the platform and the sample are subject to different configurations of the helium flow. The rotation of the platform may generate additional heat. All these factors contribute to uncertainties in the sample temperature.

In order to achieve the necessary accuracy of the anisotropy measurements in the limit $T\rightarrow T_c$, besides the ordinary PPMS-9 cryostat with rotating sample platform, we implement a home-made system devoid of the above mentioned disadvantages  as shown and explained in {Section I of Supplemental Material\cite{Suppl}}. In this system magnetic field is rotated while the sample is placed in liquid helium. The temperature is stabilized with 0.1 mK precision. 
The magnetic field was varied from positive to negative values, the data {was} symmetrized with respect to zero field to correct the Hall probe offset.

In order to ensure the selection of the single block sample, we used two indicators (see Fig.~\ref{Fig::results} a and b): {(i) a perfect figure-eight curve of the resistivity versus polar angle in a wide range of temperatures and magnetic fields and (ii) a narrow superconducting transition width (0.1-0.15 K). We choose two directions of the in-plane magnetic field corresponding to minimum ($\theta=0$) and maximum ($\theta=\pi/2$)} values of $H_{c2}$ and measure the dependence $R(H)$ at different temperatures (see Fig.~\ref{Fig::results} c and d). We calculate $H_{c2}$ using different criteria $R(H_{c2}) = \beta \cdot R_n$, where $0 < \beta < 1$ and $R_n$ is the resistivity in the normal state, as shown in Figs.~\ref{Fig::results} c and d. 

\begin{figure}[t]
\center{\includegraphics[width=0.95\linewidth]{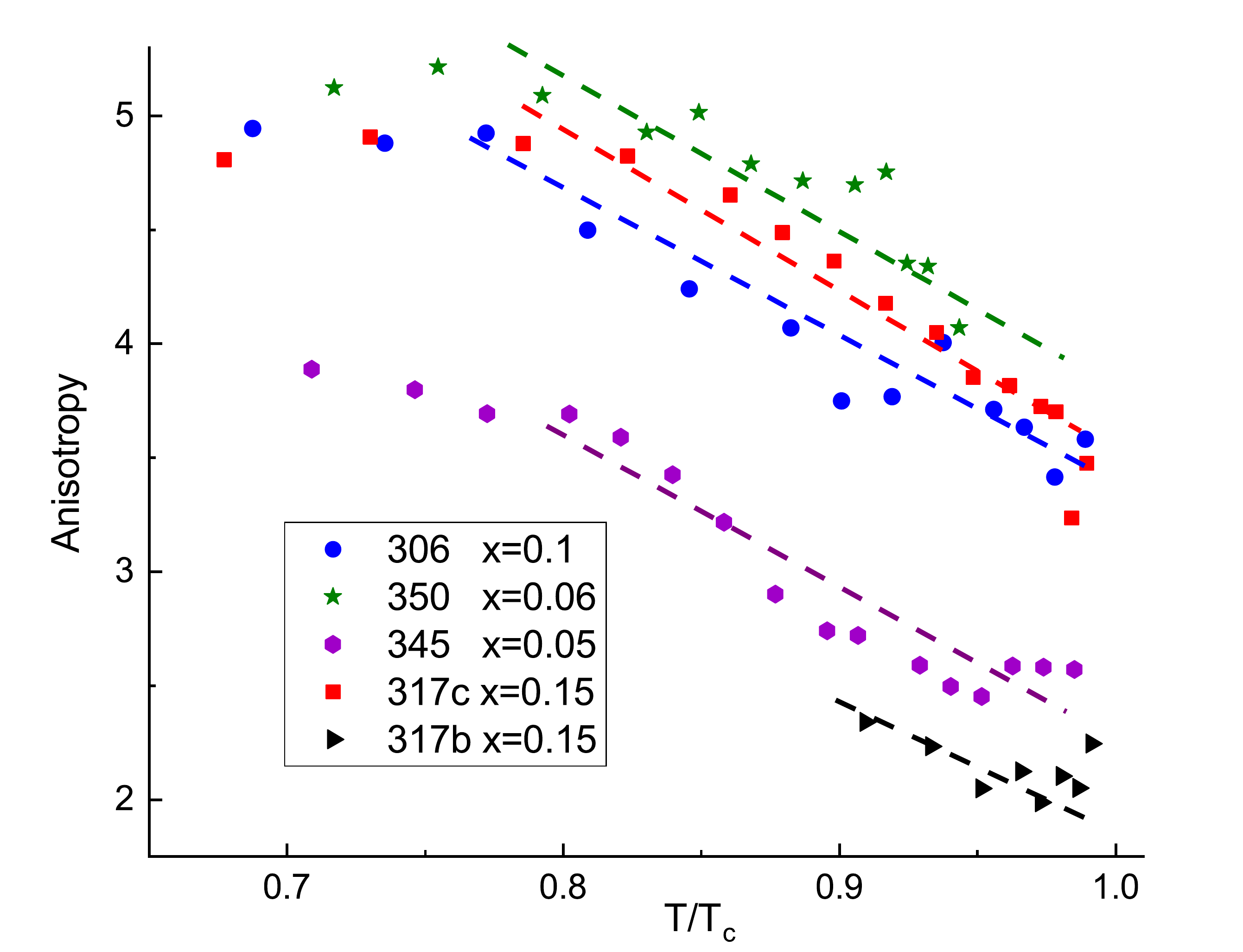}}
\caption{Statistics of the temperature dependence of the anisotropy parameter (50\% criterion) collected for different samples. {Dashed lines serve as a guide to the eye.}}
%%%%%%%%%%%%%%%%%%%%%%%%%%%%%%%%%%%%%%%%%%%%%%%%%%%%%%%%%%
\label{Fig::all_samples}
%%%%%%%%%%%%%%%%%%%%%%%%%%%%%%%%%%%%%%%%%%%%%%%%%%%%%%%%%%%
\end{figure}

Our main result, the dependence of the anisotropy of the in-plane upper critical field $An=H_{c2}(\theta=\pi/2)/H_{c2}(\theta=0)$ on $T$, is shown in Fig.~\ref{Fig::results} e. The anisotropy decreases monotonically as the temperature approaches $T_c$. This result is the same for different criteria of the calculation of $H_{c2}$ and is reproduced in different samples, see Fig. \ref{Fig::all_samples}. {Importantly, using $\beta=0.02$ criterion, that should be close to thermodynamical one, although much increases the noise of $An$ data, nevertheless reproduces the decrease of $An$ with temperature.} {We see that the anisotropy of the in-plane critical magnetic field is practically independent of $T$  below 0.8$T_c$. {Although we did not perform measurements below 1.8K, the results of Ref.\cite{Pan2016} (Fig. 3d) {clearly indicate that the anisotropy remains almost constant from 0.3K to 2K.}} At higher temperatures $An(T)$ systematically decreases, e.g. for sample 317c $An$ drops from $\approx 5$ at $T=2.3$~K to $\approx 3.5$ at $T=T_c\approx 2.7$~K (see dashed lines in Fig.~\ref{Fig::all_samples}). The anisotropy does not show any signatures of a singularity or critical behavior.}
 
{\bf Discussion.} The value of $H_{c2}$ can be calculated using a standard GL approach generalized in Ref.~\onlinecite{Venderbos2016} for the case of the vector structure of the superconducting order parameter $\mathbf{\eta}=(\eta_1,\eta_2)$.  For the readers convenience some details of the derivation are given in {Section II of Supplemental Material\cite{Suppl}}.
Here we present only the results. 

In the experiment, the applied magnetic field is directed in the basal plane, $\mathbf{H}=H(\cos{\theta},\sin{\theta})$. The anisotropy of the in-plane critical field $H_{c2}$ results from a rotational symmetry breaking. It is commonly accepted that in the case of doped Bi$_2$Se$_3$ this breaking occurs due to the strain of the crystals~\cite{Venderbos2016,KuntsevichPRB2019}. The strain can arise either in the process of the crystal growth or spontaneously~\cite{Kuntsevich2018,KuntsevichPRB2019,kapranov}. In the GL approach the effect of the strain is characterized by the value $\delta=g_N(u_{xx}-u_{yy})$, where $g_N$ is a proper GL coefficient and $u_{ik}$ are the components of the strain tensor (see {Section II of Supplemental Material\cite{Suppl}} for more details). The sign of $\delta$ depends on the choice of the axis and further we assume that $\delta>0$.   

\begin{figure}[t]
\center{\includegraphics[width=0.95\linewidth]{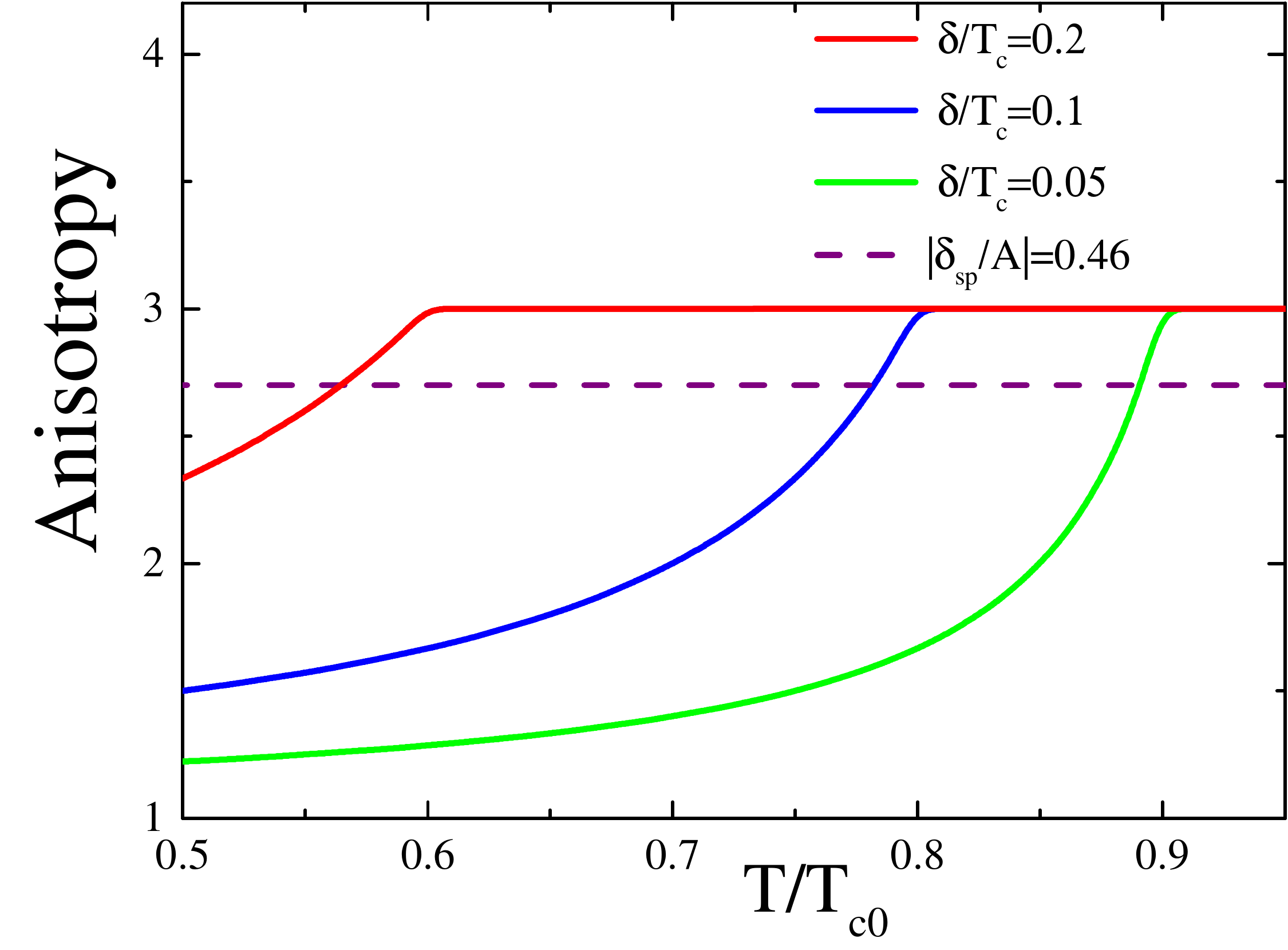}}
\caption{Anisotropy of the 
{upper critical field} %second critical field typo (iii)
vs temperature calculated from Ginzburg-Landau equations for different values of the initial strain (solid lines) and in case of the spontaneous strain $\delta_{sp}\propto 1-T/T_c$ (dashed line) for $J_4/J_1=5/4$.}
\label{Fig::phases_mu_r}
\end{figure}

For each component of the order parameter we can formally define the 
{upper critical field} %second critical field typo(iii)
$H^i_{c2}$ that corresponds to vanishing $\eta_i$, $i=1,2$. Following the procedure developed in Ref.~\onlinecite{Venderbos2016}, we derive  
%\begin{equation}\label{1}
%\!\!\!\!H_{c2}^1\!=\!\frac{-c(A+\delta)}{2e\sqrt{\!J_3\!\left[J_1\!-\!J_4\alpha(\theta)\right]}},\,\, \\
%H_{c2}^2\!=\!\frac{-c(A-\delta)}{2e\sqrt{\!J_3\!\left[J_1\!+\!J_4\alpha(\theta)\right]}}.
%\end{equation}
%Here $\alpha(\theta)=1$ if $\theta=0$ and $\alpha(\theta)=-1$ if $\theta=\pi/2$. Note that $A<0$ and $J_1,J_3,J_4>0$. We obtain from Eqs.~\eqref{1} 
\begin{eqnarray}\label{AJAd}
\nonumber
\frac{H_{c2}^2}{H_{c2}^1}(\theta=0)&=&\frac{A_\delta}{A_J},\quad\frac{H_{c2}^2}{H_{c2}^1}(\theta=\pi/2)=A_\delta A_J,\\
A_J&=&\sqrt{\frac{J_1+J_4}{J_1-J_4}}, \quad A_{\delta}=\frac{A-\delta}{A+\delta}
\end{eqnarray}
The parameters $A_J>1$ and $A_{\delta}>1$ since $\delta>0$. Thus, we can write the final formula for the anisotropy of the upper critical field in the form
\begin{equation}\label{GL_result}
An=\frac{H^2_{c2}(\theta=\pi/2)}{\max(H^1_{c2},H^2_{c2})(\theta=0)}=\min{(A_J,A_{\delta})}.
\end{equation}
The obtained formula for $An(T)$ is illustrated in Fig.~\ref{Fig::phases_mu_r}. Here we assume that $\delta$ and $J_{1,4}$ are independent of $T$. The anisotropy rapidly increases with the approach of temperature to $T^*=T_c\left[1-|\delta/A(0)|\right]$ when $A(T)+\delta\rightarrow 0$, and $An(T)$ becomes constant $A_J$ at $T>T^*$. 

{The experimentally obtained dependence of $An$ on $T$, Figs.~\ref{Fig::results} and \ref{Fig::all_samples}, evidently differs from the GL prediction,  Fig.~\ref{Fig::phases_mu_r}. In derivation of Eq.~\eqref{GL_result} we assume that the strain is external. If the strain is spontaneous, then $\delta\propto 1-T/T_c$~\cite{kapranov} and $An$ is independent of $T$, which also contradicts the experiment. In general, the coupling constant $g_N$ between the order parameter and the strain can vanish at $T=T_c$. In this case $An$ is temperature independent if the strain is also independent of $T$ or $An(T_c)=1$ if the strain is spontaneous, which is also far from the results of our measurements. }

{We see that the GL theory can not describe the drop of the anisotropy of the {upper} critical field with temperature above 0.8$T_c$, shown by dashed lines in Fig.~\ref{Fig::all_samples}. One could suppose that the GL coefficients $J_{1,4}$ depends on $T$. In the case $An=A_J$, Eq.~\eqref{AJAd}, the increase of difference $J_1-J_4$ by the factor four is enough for the decrease of $An$ from 5 to 2.5. However, such a strong variation of the GL coefficients in the temperature range about 0.4~K below $T_c$ does not seem to have any physical reason and does not arise in a direct calculations of the gradient GL coefficients~\cite{Zyuzin2017}. The second mechanism of the anisotropy drop near $T_c$ could be a weak non-uniformity of the sample, which becomes important close to $T_c$. The width of the superconducting transition for our samples is about 0.1~K at zero magnetic field and increases with the increase of the applied magnetic field, see Fig.~\ref{Fig::results} b and c. The resistance near the transition is mainly controlled by the volume fraction $p_n$ of the material in the normal state, which increases with the increase of the magnetic field. However, in the first approximation, this value scales as $p_n=p_n[H/H_{c2}(\theta)]$ which means that the non-uniformity does not strongly affect the anisotropy of $H_{c2}$ in the framework of the GL approach.   

Note that the drop of $H_{c2}$ anisotropy with approach to $T_c$ had been observed in the two-band superconductor MgB$_2$~\cite{PhysRevB.71.012504}. In Refs.~\cite{Golubov2003,Gurevich2007} it had been shown that the anisotropy drop occurs due to different scattering of the quasi-particles by impurities in different bands that results in different coherence lengths. In this case, the GL theory is applicable for a quantitative description of the two-band superconductors only in a narrow vicinity of $T_c$. A more detailed study is necessary to confirm or deny this explanation in case of the nematic superconductors.}

Another possible explanation of the $An(T)$ decrease with $T$ could be an influence of the superconducting fluctuations on the resistance near $T_c$. It has been shown in Ref.~\onlinecite{Hecker2018} that the fluctuation contribution to the conductivity is different for different in-plane magnetic field orientations. The fluctuations reduce the resistance along the direction of the maximal critical field stronger than along the perpendicular direction. This effect could be a reason for the decrease of the anisotropy near the critical temperature  and fail of the GL calculations. {$\Delta T(T_c)$ dependence given in {Section III of the Supplemental Material\cite{Suppl}} rules out the possible decrease of the anisotropy for $T > 0.8 T_c$ due to the broadening of superconducting transition.} 

{\bf Conclusions.} We study experimentally the temperature dependence of the anisotropy $An$ of the in-plane upper critical magnetic field in a typical topological superconductor Sr$_x$Bi$_2$Se$_3$. We observe that $An(T)$ is practically independent of $T$ below $T\approx 0.8T_c$ and decreases significantly when $T$ approaches $T_c$. Such a behavior can not be understood in the framework of the Ginzburg-Landau theory for nematic superconductivity in doped topological insulators. A revision of the applicability limits of the GL theory for  description of the superconductivity in these materials is necessary.

{\bf Acknowledgements.}The work is supported by the Russian Science Foundation (Grant No. 17-12-01544). R.S.A. acknowledges partial support from the Basis foundation. {A.Y.K. and M.I.B. acknowledge partial support from the Basic research program of HSE}.

\bibliographystyle{apsrev4-1}

\newpage

{\Large \bf Supplemental materials}
\\

In this Supplemental materials we show the experimental setup to measure the anisotropy of the upper critical field with high precision (section I), derive the formula for the anisotropy of the 
{upper critical field} within Ginzburg-Landau approach (section II) and analyse the possible effect of superconducting transition broadening with temperature on anisotropy of the upper critical field (section III).
 \section{Home-made experimental setup}
\begin{figure}[b]
\center{\includegraphics[width=0.95\linewidth]{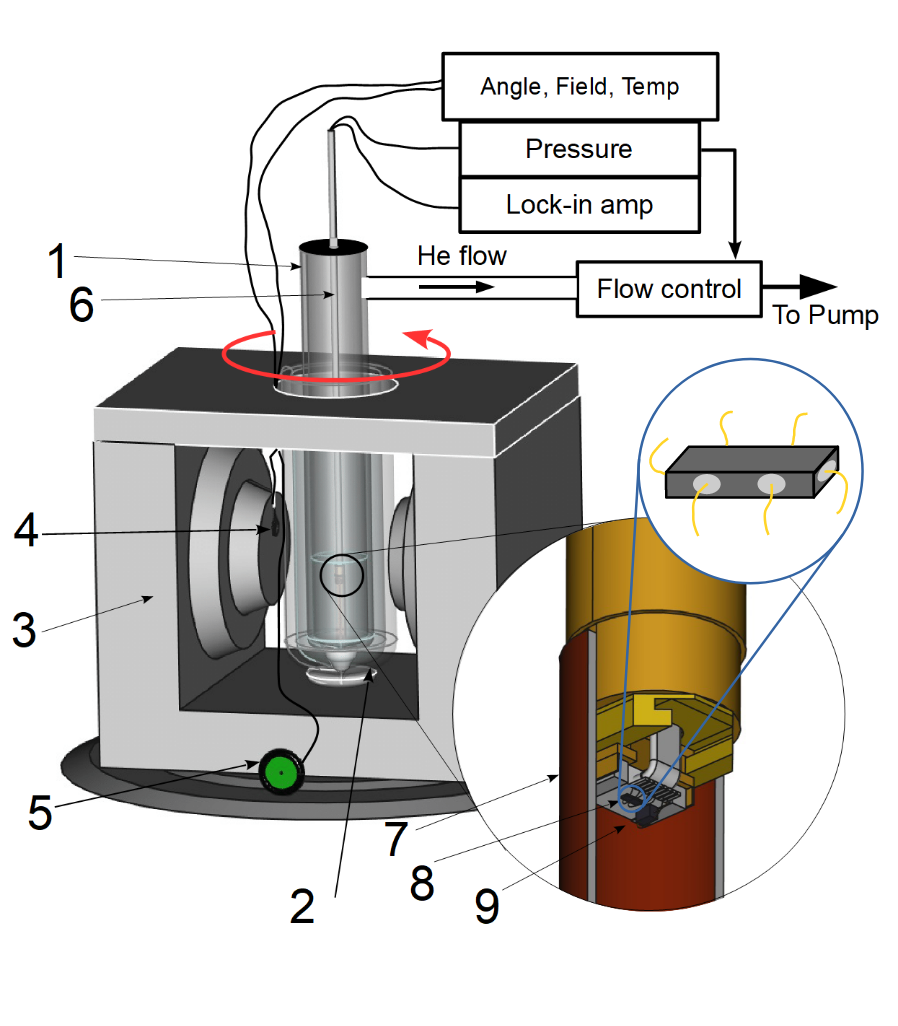}}
\caption{Schematics of the cryostat with room-temperature electromagnet.
1, 2 -- Helium and nitrogen glass dewars, 3 -- electromagnet, 4 -- Hall sensor, 5 -- angle-measuring potentiometer, 6 -- probe stick. Zoom-in shows the sample cell. 7 -- copper shield, 8 -- sample, 9 -- RuO$_2$ thermometer. {Smaller zoom-in shows the schematic view of the sample with several silver paste contacts.}}
%%%%%%%%%%%%%%%%%%%%%%%%%%%%%%%%%%%%%%%%%%%%%%%%%%%%%%%%%%
\label{Fig::scheme}
%%%%%%%%%%%%%%%%%%%%%%%%%%%%%%%%%%%%%%%%%%%%%%%%%%%%%%%%%%%
\end{figure}

 The scheme of the-home-made setup is shown in Fig. \ref{Fig::scheme}.
 The sample and thermometer are glued to a small platform at the end of the insert and are positioned inside the glass helium dewar. Highly stable (with the accuracy about 0.1~mK) liquid helium temperature is achieved by continuous pumping of it's vapor through a controllable pressure valve with a feedback from the Barathron\texttrademark\, capacitance manometer. The magnetic field up to 0.5~T is produced by an electromagnet mounted on a rotating platform. Hall sensor is located outside the outer nitrogen dewar thus it's readings are independent of He temperature. 
 
 \section{Ginzburg-Landau equations}\label{GLEq}
 The GL free energy of the nematic superconductor of Bi$_2$Se$_3$ family includes three terms~\cite{Venderbos2016}
 \begin{equation}\label{G_L}f_{\textrm{GL}}=f_{\textrm{hom}}+f_D+f_{SB}.\end{equation}
 The first term corresponding to a homogeneous state is
 \begin{eqnarray}\label{hom}
 \nonumber
 f_{\textrm{hom}}&=&A\left(|\eta_1|^2+|\eta_2|^2\right)+B_1\left(|\eta_1|^2+|\eta_2|^2\right)^2\\
 &+&B_2\left|\eta_1\eta_2^*-\eta_1^*\eta_2\right|^2,
 \end{eqnarray}
 where $\mathbf{\eta}=(\eta_1,\eta_2)$ is a two-component superconducting order parameter, $A=a (T-T_{c0})<0$, $B_1>0$, and $B_2>0$ are phenomenological GL coefficients. Here
 {$T_{c0}$ is } %$T_{c0}$is|| typo(iv)
 the critical temperature in absence of the strain.

 The gradient part of the free energy $f_D$ can be written as~\cite{Venderbos2016}
 \begin{eqnarray}\label{grad} 
 \nonumber
 f_D&=&J_1(D_i\eta_\alpha)^*D_i\eta_\alpha+J_3(D_z\eta_\alpha)^*D_z\eta_\alpha+J_4\big[|D_x\eta_1|^2 \\
 &+&|D_y\eta_2|^2\!-\!|D_x\eta_2|^2\!-\!|D_y\eta_1|^2\!+\!(D_x\eta_1)^*D_y\eta_2\\
 \nonumber
 &+&(D_y\eta_1)^*D_x\eta_2+(D_x\eta_2)^*D_y\eta_1+(D_y\eta_2)^*D_x\eta_1\big],
 \end{eqnarray}
 where $D_i=-i\partial_i-(2e/c)A_i$ is the gauge-invariant gradient ($\hbar=1$), $\mathbf{A}$ is the vector potential, $J_i$ are the GL phenomenological coefficients, and summation over repeated indices $i=x,y$ and $\alpha=1,2$ is assumed. The GL theory with the free energy $f_{\textrm{GL}}=f_{\textrm{hom}}+f_D$ describes a superconductivity with a nematic order parameter $\mathbf{Q}=(|\eta_1|^2-|\eta_2|^2,\eta_1^*\eta_2+\eta_2^*\eta_1)$. All possible directions of the nematicity $\mathbf{n}=(\cos \theta, \sin \theta)$ are equally favorable and the value of the upper critical field $H_{c2}$ is independent of the angle $\theta$ between applied magnetic field $\mathbf{H}$ and the crystal axis. 

 Strain is responsible for the third term $f_{SB}$. We assume here for simplicity that the strain is a uni-axial. In this case we have~\cite{Venderbos2016} 
 \begin{equation}\label{f_SB}
 f_{SB}=\delta\left(|\eta_1|^2-|\eta_2|^2\right).
 \end{equation}
 Here $\delta = g_N(u_{xx}-u_{yy})$, $g_N$ is the coupling constant, and $u_{ik}$ are corresponding components of the deformation tensor. As a result, the degeneracy of the nematicity direction is omitted and the in-plane critical magnetic field acquires the angle dependence, $H_{c2}(\theta)$.  

 Critical temperature is determined as $T_c=\max (T_+,T_-)$ where $a(T-T_{c0}) \pm \delta(T_{\pm})=0$. If 
 {$\delta(T_{\pm})$ }%$\delta(T_{\pm}$ typo(v)
 is finite at the transition temperature then deformation increases critical temperature $T_c=T_{c0}+|\delta/a|$ and changes type of the transition from the second to the first order. If $\delta(T_{c})=0$ vanishes at the critical temperature then there is no effect of the deformation on the critical temperature and the type of the transition. 

 \subsection{Calculations}

 To calculate the upper critical magnetic field $H_{c2}$ we minimize the free energy and use corresponding linearized GL equations for components of the order parameter. We choose the Landau calibration of the vector-potential, $\mathbf{A}=(0,0,A_z)$, in which $A_z=H(-x\sin{\theta}+y\cos{\theta})$, where $\mathbf{H}=H(\cos{\theta},\sin{\theta})$ is the in-plane applied magnetic field. In addition, we should take into account that $\eta$ is constant in the direction along the magnetic field. After straightforward algebra we obtain 
 \begin{equation}\label{eta}
 \![A\!+\!J_1 \partial_{\perp}^2\!+\!J_3D_{z}D_{z}^*\!-\!J_4\partial_{\perp}^2(\cos{2\theta} 
 \tau_z\! +\!\sin{2\theta}\tau_x)\!+\!\delta \tau_z]\eta\!=\!0,
 \end{equation}
 where $\tau_i$ are the Pauli matrices and $\partial_{\perp}$ is the component of differential operator in the $(x,y)$ plane transverse to the magnetic field direction. The function $H_{c2}(\theta)$ achieves its maximum or minimum value when $\theta=0$ or $\pi/2$ depending on the sign of the strain. Below we consider only these two cases, when Eqs.~\eqref{eta} are decoupled since $\sin{2\theta}=0$ for both $\theta=0$ and $\theta=\pi/2$. 

 We introduce position ${\hat X}=l_bD_{\perp}$ and momentum ${\hat P}=l_bD_{z}$ operators with commutation rule $[{\hat X},{\hat P}]=i$ ($\theta=0$) or $[{\hat X},{\hat P}]=-i$ ($\theta=\pi/2$), where $l_b^2=c/2eH$ is the magnetic length. We rewrite Eq.~\eqref{eta} in the form
 \begin{eqnarray}
 ({\hat P}^2&+&\omega_i^2{\hat X}^2)\eta_i=\epsilon_i \eta_i, \\
 \omega_{1}^2&=&\frac{J_1-J_4\alpha(\theta)}{J_3}, \qquad \omega_{2}^2=\frac{J_1+J_4\alpha(\theta)}{J_3}, \\
 \epsilon_1&=&-l_b^2\frac{A+\delta}{J_3}, \qquad \epsilon_2=-l_b^2\frac{A-\delta}{J_3}
 \end{eqnarray}
 Here $\alpha(\theta)=1$ if $\theta=0$ and $\alpha(\theta)=-1$ if $\theta=\pi/2$. Thus, in the considered cases, the linearized GL equations are decoupled and the components of the nematic order parameter $\eta_1$ and $\eta_2$ are independent. For each component we can formally define the 
 {upper critical field} %second critical field typo(iii)
 $H_{c2}^i$ that corresponds to the vanishing component $\eta_i$. Following a standard procedure~\cite{Venderbos2016}, we derive  
 \begin{equation}\label{hc212}
 \!\!\!\!H_{c2}^1\!=\!\frac{-c(A+\delta)}{2e\sqrt{\!J_3\!\left[J_1\!-\!J_4\alpha(\theta)\right]}},\,\, \\
 H_{c2}^2\!=\!\frac{-c(A-\delta)}{2e\sqrt{\!J_3\!\left[J_1\!+\!J_4\alpha(\theta)\right]}}.
 \end{equation}
 Note that $A<0$, $J_1,J_3,J_4>0$, and we suppose that $\delta>0$. We obtain from Eqs.~\eqref{hc212} 
 \begin{eqnarray}
 \nonumber
 \frac{H_{c2}^2}{H_{c2}^1}(\theta=0)&=&\frac{A_\delta}{A_J},\quad\frac{H_{c2}^2}{H_{c2}^1}(\theta=\pi/2)=A_\delta A_J,\\
 A_J&=&\sqrt{\frac{J_1+J_4}{J_1-J_4}}, \quad A_{\delta}=\frac{A-\delta}{A+\delta}
 \end{eqnarray}
 The parameters $A_J>1$ and $A_{\delta}>1$ since we choose $\delta>0$. Thus, we can write equation for the anisotropy of the 
 {upper critical field} %second critical field typo(iii)
 in the form
 \begin{equation}
 An=\frac{H^2_{c2}(\theta=\pi/2)}{\max(H^1_{c2},H^2_{c2})(\theta=0)}=\min{(A_J,A_{\delta})}.
 \end{equation}

 The dependence of the anisotropy of $H_{c2}$ on temperature arises due to the temperature dependence of the ratio $J_4/J_1$ and of the normalized strain $\delta/A$. In general, the temperature dependence of the ratio $J_4/J_1$ is smooth near $T_{c0}$, while $A=a(T-T_{c0})$ vanishes at $T_{c0}$ and if $\delta(T_{c0}) \neq 0$, then, the normalized strain diverges at $T=T_{c0}$, $\delta/A \propto (T-T_{c0})^{-1}$. As a result, the anisotropy increases and reaches the value $An=A_J$ at some temperature $T^*$ below $T_{c0}$, $T^*=T_{c0}-|\delta/a|$. However, the normalized strain may be finite at $T=T_{c0}$ if the coupling constant between the superconducting order parameter and strain becomes zero in the normal state. In this case the temperature dependence of $An$ is smooth near the superconducting to normal transition. Therefore, we can extract information about the coupling constant between $\eta$ and the strain from the dependence of the anisotropy of $H_{c2}$ on temperature.

{\section{Superconducting transition broadening}
To make sure that the decrease of the anisotropy for $T > 0.8 T_c$ is not caused by the broadening of $\Delta T$ in the raw data we have performed a 2d-interpolation of resistivity as a function of temperature and magnetic field for max- (fig.\ref{fig::Delta-T-evolution}(a)) and min-$H_{c2}$ (fig.\ref{fig::Delta-T-evolution}(b)) directions, respectively. 

\begin{figure}[h]
\center{\includegraphics[width=\linewidth]{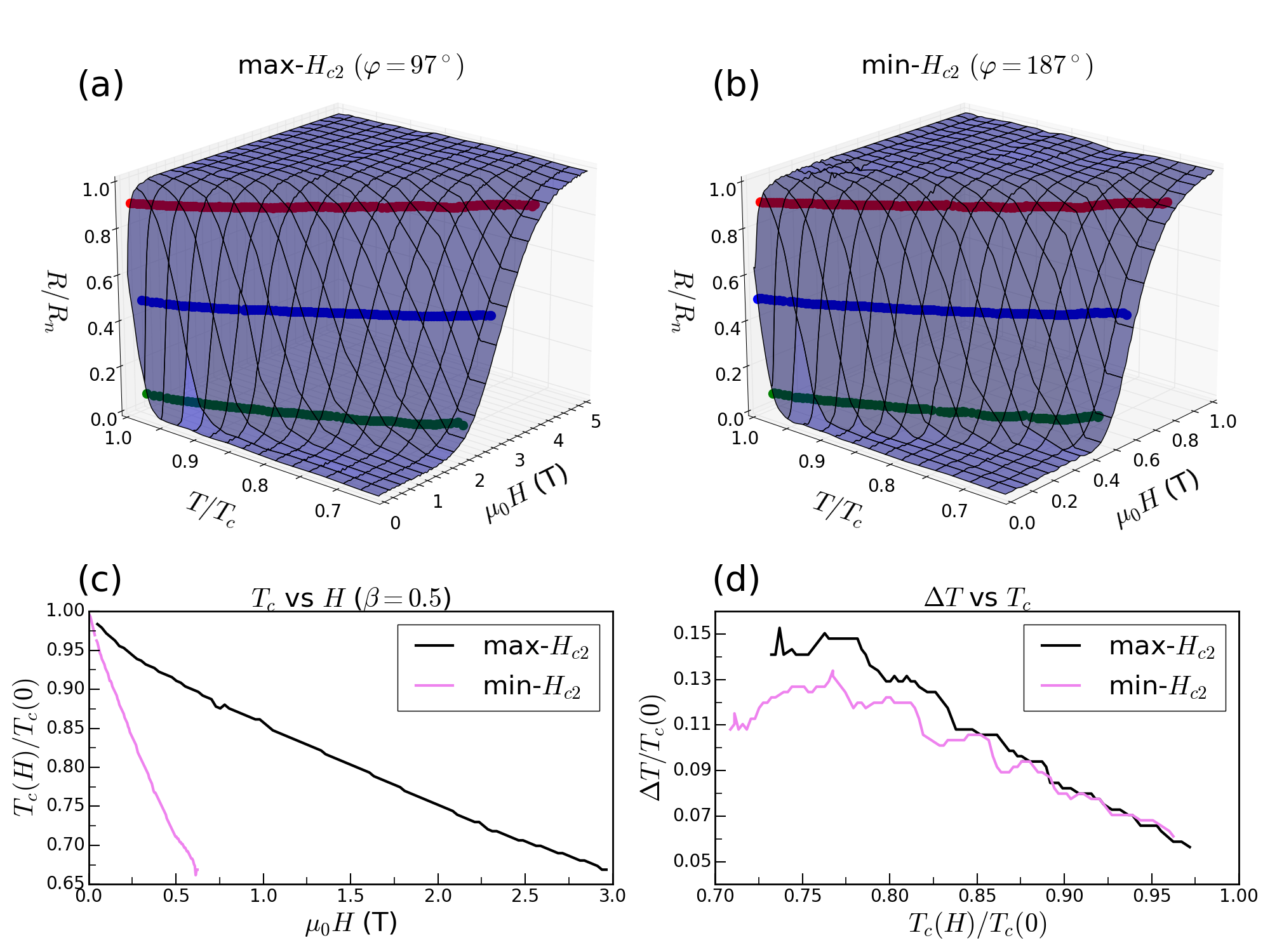}}
\caption{2d-interpolation of resistivity as a function of temperature and magnetic field for max-$H_{c2}$ (a) and min-$H_{c2}(b)$ directions. Green, blue and red lines consist of points satisfying $R(T, H) = \beta R_n$ for $\beta = 0.1,\,0.5,\,0.9$ respectively. (c) Evolution of $T_c$ for $\beta = 0.5$ as a function of $H$ for max- and min-$H_{c2}$ directions. (d) Evolution of $\Delta T$ as a function of $T_c(H)$ for max- and min-$H_{c2}$ directions.} 
\label{fig::Delta-T-evolution}
\end{figure}

First we extract from this data $T_c$ as a function of $H$ for three different criteria: 10\%, 50\%, 90\% and designate those dependencies as $T_{c}^{(1)}(H)$, $T_{c}(H)$ and $T_{c}^{(2)}(H)$, i. e.
\begin{align*}
    R\left(T_c^{(1)}(H), H\right) &= 0.1 \,R_n; \\
    R\left(\vphantom{T_c^{(1)}}T_c(H), H\right) &= 0.5 \,R_n; \\
    R\left(T_c^{(2)}(H), H\right) &= 0.9\, R_n, 
\end{align*}
where $R_n$ is the resistivity of normal state. $T_c(H)$ dependencies are shown on fig.\ref{fig::Delta-T-evolution}(c). We define $\Delta T$ as $T_c^{(2)} - T_c^{(1)}$ and plot $\Delta T$ vs $T_c$ graphs for max- and min-$H_{c2}$ directions on fig.\ref{fig::Delta-T-evolution}(d).  

It is clearly seen from our data that the transition narrows (i. e. $\Delta T$ decreases) with the increase of the temperature. Moreover, $\Delta T$ for max- and min-$H_{c2}$ directions is almost identical for $T_c(H) > 0.85\, T_c(0)$. Therefore, we do not believe that broadening of the transition could affect the anisotropy of the upper critical field.
}

\end{document}